\documentclass[aps,prb,amsmath,amssymb,amsfonts,superscriptaddress,floatfix,nofootinbib]{revtex4} 
\usepackage{amsmath,amsthm,amsfonts,amssymb}
\usepackage{graphicx}

\usepackage{color}
\numberwithin{equation}{section}
\numberwithin{figure}{section}

\newtheorem{theorem}{Theorem}

\newtheorem{lemma}{Lemma}

\newtheorem{definition}{Definition}

\newcommand{\sO}{\tilde O}
\newcommand{\be}{\begin{equation}}
\newcommand{\ee}{\end{equation}}
\newcommand{\kb}{\kappa}
\newcommand{\lb}{\lambda}

\begin{document}

\title{Local Maxima and Improved Exact Algorithm for MAX-2-SAT}

\affiliation{Station Q, Microsoft Research, Santa Barbara, CA 93106-6105, USA}
\affiliation{Quantum Architectures and Computation Group, Microsoft Research, Redmond, WA 98052, USA}
\author{Matthew B.~Hastings}

\begin{abstract}
Given a  MAX-2-SAT instance, we define a local maximum to be an assignment such that changing any single variable reduces the number of satisfied clauses.  We consider the question of the number of local maxima that an instance of MAX-2-SAT can have.
We give upper bounds in both the sparse and nonsparse case, where the sparse case means that there
is a bound $d$ on the average number of clauses involving any given variable. 
The bounds in the nonsparse case
are tight up to polylogarithmic factors, while in the sparse case the bounds are tight up to a multiplicative factor in $d$ for large $d$.
Additionally, we generalize
to the question of assignments which are maxima up to changing $k> 1$ variables simultaneously; in this case, we give
explicit constructions with large (in a sense explained below) numbers of such maxima in the sparse case.
The basic idea of the upper bound proof is to consider a random assignment to some subset of the variables and determine the probability that some fraction of the remaining variables can be fixed without considering interactions between them.
The bounded results hold in the case of weighted MAX-2-SAT as well.
Using this technique and combining with ideas from Ref.~\onlinecite{golo}, we find an algorithm for weighted MAX-2-SAT which is faster for large
$d$ than previous algorithms which use polynomial space; this algorithm does require an additional bounds on maximum weights and degree.
\end{abstract}
\maketitle
\section{Introduction}
Local search algorithms for combinatorial optimization problems
such as MAX-SAT can be trapped in local maxima and hence fail to find the global maximum.  A natural question then is: how many local maxima can an optimization problem have?
We first consider the question of assignments which are maxima when a single variable's assignment is changed,
and we find tight bounds on the number of such maxima
up to polylogarithmic factors for nonsparse MAX-2-SAT instances
and find other bounds for sparse instances (tight up to certain constants in the exponent explained later).
The methods used to prove these bounds lead to an algorithm for weighted MAX-2-SAT which is faster for high degree  instances than any previously known algorithm which uses only polynomial space (there is an algorithm of Williams which uses exponential space\cite{williams} and is exponentially faster); this algorithm requires combining these results 
with previous results of Golovnev and Kutzkov\cite{golo} and the algorithm does require some additional bounds on maximum weights and degree.

The formal definition of a local maximum will be:
\begin{definition}
\label{localmaxdef}
Given an instance of MAX-2-SAT, an assignment is a ``local maximum" if it has the property that changing the assignment to any single
variable reduces the number of satisfied clauses, while a ``global maximum" is an assignment which maximizes the number of satisfied clauses.  (We define local and global maxima for weighted MAX-2-SAT similarly, replacing the number of satisfied clauses with a sum of weights of satisfied clauses.)
\end{definition}
In section \ref{kmin}, we give a generalization of this definition to changing the assignment to $k=O(1)$ variables simultaneously.  We call these $k$-maxima and we construct instances
with large numbers of such $k$-maxima.

Note that it is clearly possible for a MAX-2-SAT instance with $N$ variables to have $2^N$ global maxima: simply take an instance with no clauses so that every assignment is a global maximum.  However, none of these global maxima are local maxima according to this definition.
The following construction\cite{csse} shows that
it is possible for a MAX-2-SAT instance to have $\Theta(N^{-1/2}) 2^{N}$ local maxima.  Assume $N$ is even.
For each pair of variables $b_i,b_j$, with $1 \leq i <j \leq N$, we have clauses $b_i \vee b_j$ and $\overline b_i \vee \overline b_j$.
There are $2 {N \choose 2} \approx N^2$ clauses in total.  For every pair $i,j$, at least one of these clauses is satisfied, and both are satisfied if $i$ is true and $j$ is false or vice-versa.  If $n$ of the variables are set to true and the remainder are false, the total number of satisfied clauses is ${N \choose 2} + n \cdot (N-n)$.  This is maximized if $n=N/2$ so that every assignment with $n=N/2$ is both a local maximum and a global maximum.  Thus, there are ${N \choose N/2} = \Theta(N^{-1/2}) 2^N$ local maxima.
Note that if we instead consider MAX-CSP with constraints that are allowed to involve an arbitrary number of variables, then by taking a single clause which is simply a parity function of the variables, we obtain an instance with $(1/2) 2^{N}$ local maxima.  While it is not hard to show that this number $(1/2) 2^N$ is optimal for a MAX-CSP instance\cite{maxcsp}, a natural question is whether
$\Theta(N^{-1/2}) 2^N$ is the maximum possible number of local maxima for a MAX-2-SAT instance.  In this paper, we prove an upper bound by ${\rm polylog}(N) N^{-1/2} 2^N$ for the number of local maxima.

We define the degree of a variable $i$ in a MAX-SAT instance to be the number of other variables $j$ such that there is a clause depending on $b_i$ and $b_j$; note that this does not depend upon the number of such clauses or the weights of such clauses.
The construction of Ref.~\onlinecite{csse} can be modified to give an instance with a large number of local maxima and with bounded degree, by taking multiple copies of the construction.
The construction of Ref.~\onlinecite{csse} has maximum degree $d=N-1$.
Consider an instance with $N$ variables, all of degree $d$, with $N$ being a multiple of $d+1$, such that the instance consists of $N/(d+1)$ decoupled copies of the construction of Ref.~\onlinecite{csse}.  This gives an instance with $N$ variables, degree $d$, and $$\Bigl(\Theta((d+1)^{-1/2}) 2^{d+1}\Bigr)^{N/(d+1)}=
2^{N\cdot(1-\frac{1}{2}\frac{\log_2(d)}{d}-O(1/d))}$$
local maxima, where here the $O(\ldots)$ notation refers to asymptotic dependence on $d$ (we use big-O notation for dependence on both $d,N$ in this paper and it should be clear by context what is intended).
In this paper, we prove an upper bound by 
$2^{N \cdot (1-\kb \log_2(\lb d)/d))}$ on the number of local maxima in the bounded degree case, for some constants $\kb,\lb>0$; since the constant $\kb<1/2$, this bound
will not immediately imply the unbounded degree bound, and we prove the two bounds separately.   Our upper bounds in the bounded degree case will hold also for bounded {\it average} degree.  This upper bound is tight up to a multiplicative factor in $d$ for large $d$: for any $\kappa'<\kappa$ for all sufficiently large $d$, the upper bound on the possible number of local maxima for degree $d$ is smaller than the lower bound for degree $\kappa' d$ which follows from the above construction (the need to take $\kappa'<\kappa$ and to take $d$ large is due to terms $N\cdot O(1/d)$ in the exponent).

The bounds in both the bounded and unbounded degree case rely on the same basic idea.  We find a subset $T$ of the variables so that interactions between pairs of variables in $T$ are small compared to interactions between variables in $T$ with those outside $T$.  Then, we show that, for a random assignment of variables not in $T$, one often finds that for many of the variables in $T$, the value of that variable at a local or global maximum can be fixed by the assignment to variables outside $T$.
The simplest version of this argument gives a 
bound of $2^{N\cdot (1-1/(d+1))}$ on the number of local maxima in the case of bounded maximum degree as follows: construct a graph whose vertices correspond to variables and with an edge between variables if they are both in a clause.  This graph can be $(d+1)$-colored and at least one color has at least $N/(d+1)$ variables.  Let $T$ be the set of variables with that color so that there is no interaction between variables in $T$.
Then, there is at most one local maximum for every assignment to variables outside $T$, so that
there are at most $2^{N\cdot (1-1/(d+1))}$ local maxima.  The stronger bound in the present paper involves choosing a larger set $T$ so that the interaction between variables in $T$ may not vanish; this require a more complicated probabilistic estimate.

This kind of bound of local maxima naturally leads to an algorithm to find a global maximum: iterate over assignments to variables outside $T$.  Then, for each such assignment, for many of the variables in $T$, one can determine the optimal assignment to that variable considering only its interaction to variables outside $T$.  We will thus obtain an algorithm which takes time
$\sO(2^{N\cdot (1-\kb \log_2(\lb d)/d))})$.

Previous algorithms for MAX-2-SAT include an algorithm taking time $\sO(2^{\omega N/3})$ but using exponential space\cite{williams} where $\omega$ is the matrix multiplication exponent.  Among algorithms using polynomial space\cite{prev1,prev2,golo}, the fastest\cite{golo} for large $d$ takes time $\sO(2^{N\cdot(1-\alpha\ln(d)/d)})$ for any $\alpha<1$, while others took time $\sO(2^{N\cdot(1-{\rm const.}/d)})$ for various constants.
This algorithm\cite{golo} is faster for large $d$ than the algorithm given in the above paragraph.  However, we show how to combine ideas of the two algorithms to obtain a faster algorithm for large $d$, subject to some additional bounds on maximum degree and weights explained later.

Some notation: if not otherwise stated, logarithms are always to base $2$.  We use ${\rm const.}$ to denote various positive numerical constants throughout.  When we need a more specific value of a constant, we introduce a symbol such as $c,\kb,\lb,\ldots$.  We use $\Pr(\ldots)$ to denote probabilities.  We use $\sO(\ldots)$ to denote asymptotic behavior up to a polylogarithm of the argument; when the argument is an exponential in $N$, this will be a polynomial in $N$.
Given a set of variables $T$, we use $\overline T$ to denote the variables not in that set.  As a final point of terminology, we will translate (for notational convenience) both weighted and unweighted MAX-2-SAT instances into ``Ising instances" and we will almost exclusively use the notation of Ising instances later in this paper.  Whenever we refer to MAX-2-SAT instances, we will be considered with assignments that are maxima, but for Ising instances these will become minima.

We begin in section \ref{defn} with definitions and we introduce the notion of an ``effective Hamiltonian", which is an optimization problem for some subset of variables given an assignment to the others.  Then, in section \ref{kmin}, we introduce the notion of $k$-minima and $k$-maxima and give constructions with large numbers of such minima;  section \ref{kmin} is independent of the following sections of the paper and can be skipped if desired.  The main result in this section is theorem \ref{kminmany}. 
 We show that there is a constant $c>0$ such that for any $f,l$ there is an Ising instance which has degree
$d=f(l-1)$
and which has at least
$2^{N\cdot(1-\frac{f \log(f\sqrt{l}/c)}{l})}$ global minima, all of which have Hamming distance at least $2^f$ from each other.
This result implies that one can obtain instances with many such disconnected minima, where there are ``many" minima in that
the term $1-\frac{f \log(f\sqrt{l}/c)}{l}$ in the exponent is only slightly smaller than $N$ once $d$ is large.
One could then add additional terms to the objective function to make one of those minima the global minimum while raising all the others in energy slightly, giving an instance with many local minima which are all separated by Hamming distance $2^f$ and with a unique global minimum.

 Then, in section \ref{nsp}, we give upper bounds on the number of local minima of an Ising instance without any degree bound.  The main result in this section is theorem \ref{nspthm}.
Then, in section \ref{spc}, we give upper bounds on the number of local minima of such an instance assuming a degree bound.  The main result in this section is theorem \ref{badcase}.
In section \ref{secrandvar} we prove a technical lemma on sums of random variables which we use in sections \ref{nsp},\ref{spc}; this technical lemma is only needed because we are concerned with the weighted case and we must consider arbitrary weights; with bounds on the weights we would not need this lemma  as we could instead use a Berry-Esseen theorem.
Finally, in section \ref{combine} we show how to combine the ideas here with those in Ref.~\onlinecite{golo} to obtain a faster algorithm.

\section{MAX-2-SAT Definitions and Effective Hamiltonians}
\label{defn}
\subsection{Problem Definitions}
We consider a MAX-2-SAT instance, with variables $b_i$ for $i=1,\ldots,N$ taking values true or false.  We re-write the instance as an Ising model to make the notation simpler, setting $S_i=+1$ if $b_i$ is true and $S_i=-1$ if $b_i$ is false.
Each 2-SAT clause can be written as a sum of terms which are quadratic and linear in these variables.
For example, clause $b_i \vee b_j$ is true if $b_i$ is true or $b_j$ is true.  So, the clause is true if
$1-(1-S_i)(1-S_j)/4$ is equal to $1$ and is $0$ otherwise.
The negation of a variable (replacing $b_i$ by $\overline b_i$) corresponds to replacing $S_i$ with $-S_i$.
Note that $1-(1-S_i)(1-S_j)/4=3/4+S_i/4+S_j/4-S_i S_j/4$.
So, given $C$ clauses, we can express the number of violated clauses as an expression of the form
\be
\label{Ising}
H=-\frac{3}{4}C+\frac{1}{4}\sum_i h_i S_i + \frac{1}{4}\sum_{i<j} J_{ij} S_i S_j,
\ee
where $h_i,J_{ij}$ are integers.  We will set $J_{ij}=J_{ji}$ and $J_{ii}=0$.

We refer to an $H$ such as in Eq.~(\ref{Ising}) as a ``Hamiltonian".  We drop constant terms such as $(3/4)C$ throughout. 

We refer to the problem of minimizing an expression such as $H$ over assignments as the Ising problem (as we explain in the next few paragraphs, we will in fact allow $h_i,J_{ij}$ to be arbitrary reals for the Ising problem but we will assume that certain operations involving $h_i,J_{ij}$ can be done in polynomial time).  
Below, we construct algorithms to solve the Ising problem and bounds on local minima for the Ising problem; this implies corresponding algorithms
and bounds for MAX-2-SAT and for weighted MAX-2-SAT (in the case of weighted MAX-2-SAT, the $h_i,J_{ij}$ need not be integers if the weights are not integers).

We will allow $h_i,J_{ij}$ to be arbitrary real numbers in what follows.  However, we will assume that all the arithmetic we do below (adding and comparing quantities $h_i,J_{ij}$) can be done in polynomial time in $N$.

\subsection{Effective Hamiltonian} 
We introduce a notion of an effective Hamiltonian, as follows.
\begin{definition}
Let $V$ denote the set of variables.  Let $T\subset V$.
Let $\overline T\equiv V \setminus T$.
Given an assignment $A$ to all variables in $\overline T$, we define an effective Hamiltonian $H^{eff}$ on the remaining variables as
\be
H^{eff}=\frac{1}{4}\sum_{i \in T} h^{eff}_i S_i + \frac{1}{4}\sum_{i,j \in T, i<j} J_{ij} S_i S_j,
\ee
where
\be
h^{eff}_i=h_i+\sum_{j \not \in  T} J_{ij} S_j.
\ee
\end{definition}
Then, given any local minimum of $H$, this local minimum determines some assignment $A$ to variables in $\overline T$ and some assignment $B$ to variables in $T$.  The assignment $A$ determines an effective Hamiltonian $H^{eff}$.  Then, the assignment $B$ must be a local minimum of $H^{eff}$.
Further, if we have an algorithm to determine a global minimum of $H^{eff}$ for any assignment $A$ to variables in $\overline T$, and given that that algorithm takes time $t$, we can find a global minimum of $H$ in time $\sO(2^{N-|T|} t)$ by iterating over assignments to variables in $\overline T$ and finding the minimum of $H^{eff}$ for each assignment.

We will  consider some special cases.
\begin{lemma}
Suppose that $H^{eff}$ has the property that $J_{ij}=0$ for all $i,j  \in T$.  Then, for every $i \in T$ such that $h^{eff}_i\neq 0$, every global minimum of $H^{eff}$
has $S_i=-{\rm sign}(h^{eff}_i)$.  
The variables $i \in T$ with $h^{eff}_i=0$ can be chosen arbitrarily at a global minimum.

There is at most one local minimum; such a local minimum exists only if there are no $i$ with $h^{eff}_i=0$; in this case, the local minimum is the unique global minimum.
 \begin{proof}
 Immediate.
 \end{proof}
 \end{lemma}
 Another special case we consider is where $J_{ij}$ may be non-vanishing for pairs $i,j \in T$, but for many $i$ we have that $|h^{eff}_i|\geq \sum_{j \in T} |J_{i,j}|$.  
  \begin{definition}
  Let
 \be
 h^{max}_i\equiv \sum_{j \in T} |J_{i,j}|.
 \ee
  \end{definition}
  \begin{definition}
 For $i \in T$, if
  $|h^{eff}_i|\geq h^{max}_i$ we say that $i$ is ``fixed", otherwise, $i$ is ``free".
  \end{definition}

  \begin{lemma}
  \label{specialcase}
  If $F$ is the set of free variables for $H^{eff}$, then $H^{eff}$ has at most $2^{|F|}$ local minima.
At every global minimum or local minimum of $H^{eff}$, for each $i$ which is fixed with $h^{eff}_i \neq 0$, we have $S_i=-{\rm sign}(h^{eff}_i)$.  
The fixed $i$ with $h^{eff}_i=0$ can be chosen arbitrarily at a global minimum.
   \begin{proof}
 Immediate.
 \end{proof}
  \end{lemma}

\section{$k$-Minima and $k$-Maxima}
\label{kmin}
In this section, we give a more general definition of local maximum or minimum, which we call a $k$-maximum or $k$-minimum.  This definition allows one to change the assignment to several variables at a time, and also generalizes in a way that is appropriate to describing equilibria of certain local search algorithms.  We then give constructions of Ising instances with large numbers of $k$-minima.

\begin{definition}
Given an assignment $A$ to a MAX-2-SAT problem, such an assignment is called a ``$k$-maximum" if every assignment
differing in at least one and at most $k$ variables from assignment $A$ satisfies fewer clauses than $A$ does.
Similarly, for an Ising Hamiltonian, an assignment $A$  is called a ``$k$-minimum" if every assignment
differing in at least one and at most $k$ variables from assignment $A$ has a larger value for $H$ than $A$ does.
\end{definition}
Hence, the definition of a local maximum or local minimum above corresponds to a $1$-maximum or $1$-minimum.

Before giving the construction of instances with large numbers of $k$-minima, we give one other possible generalization of the notion of a minimum:
\begin{definition}
Given a problem instance of MAX-2-SAT or the Ising problem, and given
integer $k\geq 1$, define a graph $G$ whose vertices correspond to assignments.  There is one vertex for each assignment such that changing at most $k$
variables of that assignment does not reduce the number of satisfied clauses (in the MAX-2-SAT case) or does not increase the value of $H$ (in the Ising case).  Let there be an edge between any two vertices which are within Hamming distance $k$ of each other.
Then, we refer to the connected components of this graphs as ``$k$-basins".
\end{definition}
Note that for every $k$-minimum, the graph in the above definition will have a connected component containing just the one vertex corresponding to that minimum, so the construction below of instances with large numbers of $k$-minima will give a construction with large
numbers of $k$-basins.  Given a local search algorithm that iteratively updates an assignment changing at most $k$ variables at a time, if the algorithm only accepts the update if it does not reduce the number of satisfied clauses, then once the assignment is in a given $k$-basin, it cannot leave that basin.

The notation using Ising Hamiltonians rather than MAX-2-SAT problems can be used to slightly simplify the construction of local minima
in Ref.~\onlinecite{csse}, so we use that notation and use the Ising problem for the rest of this section.  The Hamiltonian used to construct local minima is simply
\begin{eqnarray}
\label{lmin}
H& =&\frac{1}{2}\sum_{i< j} S_i S_j \\ \nonumber
&=& \frac{1}{4}\Bigl( \sum_i S_i \Bigr)^2 + {\rm const.}
\end{eqnarray}
Hence, the minimum is at $\sum_i S_i=0$.
The local minima of this Hamiltonian are $1$-minima.
This has degree $d=N-1$.  As explained, we can take multiple copies of this construction to give instances with $N$ variables, degree $d$, and
$2^{N\cdot(1-\frac{1}{2}\frac{\log(d)}{d}-O(1/d))}$ local minima.

Now we consider $k>1$.  We begin by giving a construction which is the analogue of the ``single copy" above for which the degree scales with $N$, and then explain how to take multiple copies of this construction to have fixed degree $d$, independent of $N$.
Pick integers $f,l>0$ with $l$ even.  The construction will give $k$-minima for $k=2^f-1$.
We have $N=l^f$ variables, labelled by a vector $(x_1,\ldots,x_f)$, with $1 \leq x_a \leq l$.
These variables may be thought of as lying on the integer points contained in an $f$-dimensional hypercube $[1,l]^f$.

A ``column" $C$ will be labelled by a choice of an integer $b$ with $1 \leq b \leq f$ and a choice of integers $y_1,y_2,\ldots,y_{b-1},y_{b+1},\ldots,y_f$ with $1 \leq y_a \leq l$.
We say that a variable $i$ is in such a column $C$ if $i$ is labelled by $(x_1,\ldots,x_f)$ with $x_a=y_a$ for all $a \neq b$.
For example, in the case $f=2$, we can regard the variables as arranged in a square of size $l$-by-$l$ and a ``column" is either a row or column of this square depending on whether $b=1$ or $b=2$.
There are $n_C$ columns, with
\be
n_C=f l^{f-1}.
\ee

Then, we take the Hamiltonian
\be
\label{Href}
H=\sum_{{\rm columns} \; C} \Bigl( \sum_{i \in C} S_i-M_C \Bigr)^2,
\ee
where $M_C$ is some integer which depends upon column $C$.
The case $f=1$ with $M_C=0$ for all columns is the Hamiltonian (\ref{lmin}) up to multiplication and addition by constants.  We use this constant $M_C$ to simplify some of the proofs later.

An assignment such that $\sum_{i \in C} S_i=M_C$ for all columns $C$ will be called a ``zero energy assignment".  Every zero energy assignment is a global minimum.

First let us consider the case that $M_C=0$ for all columns $C$.
As an explicit example of a zero energy assignment in this case, for a variable labelled by vector $(x_1,\ldots,x_f)$, take $S=+1$ if $\sum_a x_a$ is even and $S=-1$ if $\sum_a x_a$ is odd.
Heuristically, one may guess that this Hamiltonian has roughly
$$2^N (\frac{c}{\sqrt{l}})^{n_C}$$
such zero energy assignments, for some constant $c>0$.  This heuristic guess is based on the following.  There are $2^N$ possible choices of variables.  In a given column $C$, the probability that such a choice gives $\sum_{i \in C} S_i=0$ is equal to $2^{-l} {l \choose l/2} \approx c/\sqrt{l}$, for some constant $c$.
Assuming such events are independent between columns, one arrives at this heuristic guess.  Of course, the events that $\sum_{i\in C} S_i=0$ are {\it not} independent for different choices of $C$ so we need a more careful analysis.
In the case that $f=2$, enumerating the number of zero energy assignments of this Hamiltonian is a well-studied question.  It is the same as enumerating the 
number of $0-1$ matrices with $l$ rows and $l$ columns such that each row and column sums to $l/2$.  It is shown in Ref.~\onlinecite{lbound} that the heuristic guess is a lower bound for the number of such assignments; more detailed estimates are in Ref.~\onlinecite{asympt}.

However, we also want to consider the case $f>2$ and the estimates for $f=2$ do not seem to straightforwardly generalize to $f>2$.  However we have:
\begin{lemma}
Let $n_{ze}$ denote the number of zero energy assignments.  If $n_{ze}>0$, then the number of global minima is equal to $n_{ze}$.
There is a constant $c>0$ such that for any $f,l$, there exists choices of $M_C$ such that 
\be
\label{nmbound}
n_{ze} \geq 2^N (\frac{c}{f\sqrt{l}})^{n_C}.
\ee
\begin{proof}
Suppose that we choose the $S_i=\pm 1$ independently at random with $S_i=+1$ with probability $1/2$, and then define $M_C=\sum_{i\in C} S_i$.  This random choice of $S_i$ defines a probability distribution of $\Pr(\vec M)$, where $\vec M$ denotes the vector of choices of $M_C$ for each  column $C$.
To prove the bound (\ref{nmbound}), we need to show that there is some $\vec M$ such that $\Pr(\vec M)\geq (\frac{c}{f\sqrt{l}})^{n_C}$.

What we will do is estimate $\sum_{\vec M} \Pr(\vec M)^2$.  Note that
$\sum_{\vec M} \Pr(\vec M)^2 \leq {\rm max}_{\vec M}\Pr(\vec M)$, so that our lower
bound on $\sum_{\vec M} \Pr(\vec M)^2$ will immediately imply a lower bound on ${\rm max}_{\vec M}\Pr(\vec M)$
(very heuristically, one may say that we are choosing $M_C$ by picking  $S_i=\pm 1$ independently at random with $S_i=+1$ with probability $1/2$, and then defining $M_C=\sum_{i\in C} S_i$).

To estimate $\sum_{\vec M} \Pr(\vec M)^2$, we need to
consider the probability that, given two random assignments, both assignments have the same resulting $\vec M$.
To label the variables in the two different assignments, we will label
$2l^f$ variables by a vector $(x_1,\ldots,x_f)$, with $1 \leq x_a \leq f$ and by an index $\sigma=1,2$, where $\sigma$
will label one of the two assignments.
 We label columns by  a choice of an integer $b$ with $1 \leq b \leq f$ and a choice of integers $y_1,y_2,\ldots,y_{b-1},y_{b+1},\ldots$ and an index $\sigma=1,2$.
We say that a variable $i$ is in such a column $C$ if $x_a=y_a$ for $a \neq b$ and the $\sigma$ index of the variable agrees with the $\sigma$ index of
the column.
Then, to compute
$\sum_{\vec M} \Pr(\vec M)^2$
we wish to compute the probability that for every pair of columns $C_1$ with $\sigma=1$ and $C_2$ with $\sigma=2$ (with $C_1,C_2$ having the same $b$ and $y_a$) we have $\sum_{i \in C_1} S_i = \sum_{i \in C_2} S_i$.
Changing the sign of all variables with $\sigma=2$, this is the same as requiring that
$\sum_{i \in C_1} S_i - \sum_{i \in C_2} S_i=0$.

We now redefine what a column means.  For the rest of the proof of this lemma, a column will be labelled by
a choice of an integer $b$ with $1 \leq b \leq f$ and a choice of integers $y_1,y_2,\ldots,y_{b-1},y_{b+1},\ldots$, without any $\sigma$ index.  We say that a variable $i$ is in such a column $C$ if $i$ is labelled by $(x_1,\ldots,x_f)$ with $x_a=y_a$ for $a \neq b$; thus, for every column there are $2l$ variables in that column.
So, we wish to estimate the probability that $\sum_{i \in C} S_i=0$ for all columns $C$.

We can express this as an integral:
\be
\int_{[0,2\pi]^{n_C}}
\Bigl( \prod_i \cos(\sum_{C \ni i} \theta_C) \Bigr)
\Bigl( \prod_C \frac{{\rm d}\theta_C}{2\pi}\Bigr) 
=
\int_{[0,2\pi]^{n_C}}
\Bigl( \prod_{i \, {\rm s.t.} \, \sigma=1} \cos(\sum_{C \ni i} \theta_C)^2 \Bigr)
\Bigl( \prod_C \frac{{\rm d}\theta_C}{2\pi}\Bigr) 
\ee
where $\theta_C$ is integrated over from $0$ to $2\pi$ for each column $C$.  The product over $i$ in the left-hand side of this equation is over all variables $i$; using the fact that
for each vector $(x_1,\ldots,x_f)$ there are two variables labelled by that vector, with $\sigma=1,2$, we re-write the integral as in the right-hand side of the equation where we take the
product only over variables with $\sigma=1$ but we square the cosine.

A similar integral can be used to express the probability in the original problem (i.e., without the $\sigma$ index) that we have a given $\vec M$.
However, the reason we have taken this $\sigma$ index is that the cosine term is squared so that now the integral is over a positive function.
This makes it easier to lower bound the integral.

Restricting to the region of the integral with $\theta_C \leq 1/(f\sqrt{l})$, the sum $\sum_{C \ni i} \theta_C$ is bounded
by $1/\sqrt{l}$ in every case, so that $\cos(\sum_{C \ni i} \theta_C)^2$ is lower bounded by $1-1/(2l)$.  Hence,
the integrand is lower bounded by $(2\pi)^{-N_C}(1-1/(2l))^N \geq (2\pi)^{-N_C} \times {\rm const.}^{N/l}={\rm const.}^{N_C}$, for some positive constants.
The volume of this integration domain is $(f\sqrt{l})^{-N_C}$.  So, the result follows.
\end{proof}
\end{lemma}
Remark: we expect that the factor of $f$ can be removed from Eq.~(\ref{nmbound}) by a more careful estimate of the integral.
More strongly, we conjecture that a similar lower bound holds for $n_{ze}$ in the case that $M_C=0$ for all $C$.

Next, we show that
\begin{lemma}
For any choice of $M_C$ with $n_{ze}>0$, all global minima are $2^f-1$ minima. 
\begin{proof}
We must show that for every zero energy assignment, there is no other zero energy assignment within Hamming distance less than $2^f$.
We prove this inductively on $f$.
To prove the case $f=1$, consider any assignment $A$ which is a zero energy assignment.  Any assignment $B$ with Hamming distance $1$ from $A$ must have sum $\sum_i S_i$ which is either $M_C+2$ or $M_C-2$, depending on whether one changes a single $S_i$ from $-1$ to $+1$ or $+1$ to $-1$.

Now we give the induction step.  Assume the result holds for an $(f-1)$-dimensional hypercube.  We now prove it for the $f$-dimensional hypercube.
We re-write the Hamiltonian as
\begin{eqnarray}
\label{Hrefsplit}
H=\sum_{1 \leq y \leq l} H_0(y)+H_1,
\end{eqnarray}
where
\be
H_0(y)=\sum_{{\rm columns} \; C\, {\rm s.t.} \; 1 \leq b < f \, {\rm and} \, y_f=y} \; \Bigl( \sum_{i \in C} S_i-M_C \Bigr)^2,
\ee
\be
H_1=\sum_{{\rm columns} \; C {\rm s.t.} \; b = f} \Bigl( \sum_{i \in C} S_i-M_C \Bigr)^2.
\ee
That is, the columns such that $b<f$ are in the sum $\sum_y H_0(y)$ with $y=y_f$, while the columns with $b=f$ are in the sum in $H_1$.
Let assignment $A$ be a zero energy assignment.  Suppose that assignment $B$ differs from
assignment $A$ in some variable labelled by $(X_1,\ldots,X_f)$ and suppose that $B$ is a zero energy assignment also.  
Then, $A,B$ are both zero energy assignments for $H_0(X_f)$ and so by the induction hypothesis, the assignments $A,B$ differ in
at least $2^{f-1}$ variables such the label of the variable $(x_1,\ldots,x_f)$ has $x_f=X_f$.

Since $B$ is also a zero energy assignment of $H_1$, there must be some other variable
labelled by $(X_1,\ldots,X_{f-1},Z_f)$ with $Z_f\neq X_f$ such that $A,B$ differ in that variable.
Then, since $A,B$ are both zero energy assignments of $H_0(Z_f)$, again by the induction hypothesis,
  the assignments $A,B$ differ in
at least $2^{f-1}$ variables such the label of the variable $(x_1,\ldots,x_f)$ has $x_f=Z_f$.
Hence, $A,B$ differ in at least $2^f$ variables.
\end{proof}
\end{lemma}

Hence, we arrive at:
\begin{lemma}
There is a constant $c>0$ such that for any $f,l$,
there exists choices of $M_C$ such that the Hamiltonian (\ref{Href}) has at least $2^N (\frac{c}{f\sqrt{l}})^{n_C}$
global minima, all of which are $2^f-1$ minima.
\end{lemma}

So, by considering multiple copies of the above instance, using that $n_C=fN/l$ in the above construction, we find that:
\begin{theorem}
\label{kminmany}
There is a constant $c>0$ such that for any $f,l$ there is a Hamiltonian which has degree
$d=f(l-1)$
and which has at least
$$2^{N\cdot(1-\frac{f \log(f\sqrt{l}/c)}{l})}$$ global minima, all of which are $2^f-1$ minima.
\end{theorem}

\section{Non-Sparse Case}
\label{nsp}
In this section we give an upper bound on the number of local maxima for the case with no degree bound.
We first need a technical lemma, upper bounding the probability that a weighted sum of a large number of Bernoulli
random variables will fall within some interval.
We remark that if all the weights are the same, then
the desired result would follow from the Berry-Esseen theorem: we would have many random variables, all with bounded second moments (and vanishing first and third moments) and so the distribution would converge to a Gaussian up to $1/\sqrt{n}$ errors in cumulative distribution function.  However, since we allow arbitrary weights, the sum may be far from
a Gaussian and a separate proof is needed.
\begin{lemma}
\label{lemmarandvar}
Let $\sigma_i$ for $i=1,\ldots,m$ be independent random variables, uniformly chosen $\pm 1$.
Let $\Sigma=\sum_i a_i \sigma_i$.  Assume that there are at least $n$ different values of $i$ such that
with $|a_i|\geq a^{min}$, for some $a^{min}\leq \delta$.
Then,
\be
{\rm max}_h {\rm Pr}(|\Sigma+h|\leq \delta) \leq {\rm const.} \times \frac{\delta/a^{min}}{\sqrt{n}}.
\ee
\begin{proof}
This lemma is a corollary of lemma \ref{lemmarandvar2} proven in section \ref{secrandvar}.
\end{proof}
\end{lemma}

Now, we prove
\begin{theorem}
\label{nspthm}
Consider an Ising instance $H$ on $N$ variables, with $J_{ij},h_i$ arbitrary.  Then, there are at most
${\rm polylog}(N) N^{-1/2} 2^N$ local maxima.
\begin{proof}
We first construct a set $T$ of variables that are weakly coupled to each other and are at least as strongly coupled to many variables in $\overline T$, where the
strength of the coupling between two variables $i,j$ is $|J_{ij}|$.

We will pick a quantity $\epsilon$ later, with $\epsilon$ proportional to $\log(N)/N$.
Let $T_0$ be a randomly chosen set of variables with $|T_0|=\lfloor \epsilon N \rfloor$.
We will then label variables in this set $T_0$ as ``good" or ``bad".
A variable $i$ is ``good" if there are at least
$(1/2) \lfloor\epsilon^{-1}\rfloor$ variables $j \not \in T_0$ such that
$|J_{ij}| \geq {\rm max}_{k \in T_0} |J_{ik}|$.
Otherwise, $i$ is ``bad".
Colloquially, if $i$ is good, then there are at least $(1/2)\lfloor\epsilon^{-1}\rfloor$ variables not in $T_0$ which are at least as strongly coupled to $i$
as any variable in $T_0$ is.

Let us estimate the probability that for a random choice of $T_0$ that a randomly chosen variable $i$ in $T_0$ is bad.
This probability attains its maximum in the case that all $J_{ij}$ differ in absolute value for different choices of $j \neq i$.
In this case, we need to estimate the probability that given a set of $N-1$ elements all differing in magnitude, with $|T_0|-1$ elements chosen at random from this set, we choose at least one of the $(1/2)\lfloor\epsilon^{-1}\rfloor$ largest elements (i.e., that $T_0$ contains a $j$ such that $|J_{ij}|$ is one of the $(1/2)\lfloor\epsilon^{-1}\rfloor$ largest possible).  This probability is not hard to compute exactly, but we give instead a simple estimate.
The probability that any given one of these largest elements is chosen is $\leq \epsilon$.  Hence, the average number chosen is $\leq 1/2$ and so that probability that $i$ is bad is at most $1/2$.

 In case some $J_{ij}$ have the same absolute value, one can arbitrarily choose a set of $(1/2)\lfloor\epsilon^{-1}\rfloor$
distinct $j$ such that $|J_{ij}|$ for each $j$ in this set is at least as large as $|J_{ik}|$ for all $k$ not in this set, and then estimate the probability that one of the elements of this set is chosen to upper bound the probability that $i$ is bad in the same way.

Hence, for a random choice of $T_0$, the average number of good variables is at least $(1/2)\lfloor \epsilon N \rfloor$, and so
there must be some choice of $T_0$ such that there are at least $(1/2)\lfloor \epsilon N \rfloor$ good variables.  Choose $T$ to be the set of good variables for that choice of $T_0$, so that $|T| \geq (1/2)\lfloor \epsilon N \rfloor$.

Now, choose a random assignment to all variables in $\overline T$.  
Given such an assignment, we compute the effective Hamiltonian $H^{eff}$ for variables in $T$.
Recall that if
  $|h^{eff}_i|\geq  h^{max}_i=\sum_{j\not \in T} |J_{i,j}|$ we say that $i$ is ``fixed", otherwise, $i$ is free.
 We now consider the probability that a given variable $i \in T$ is free.
 We will apply lemma \ref{lemmarandvar} as follows.
 Let $a^{min}={\rm max}_{j \in T}|J_{ij}|$.  We can assume that $a^{min}>0$, otherwise $i$ is trivially fixed.
 Then, $h^{max}_i\leq a^{min}\epsilon N$ as $|T|\leq \epsilon N$,
 and $h^{eff}_i=h_i+\sum_{j \not\in  T} J_{ij} S_j$.  So, $\Pr(|h^{eff}_i|>h^{max}_i)\leq {\rm max}_h \Pr(|h+\sum_{j \not\in  T} J_{ij} S_j| \leq a^{min} \epsilon N)$.  There are at least $(1/2)\lfloor\epsilon^{-1}\rfloor$ choices of $j\not \in T$ such that $|J_{ij}|\geq a^{min}$,
 so by lemma \ref{lemmarandvar}, the probability that $i$ is free is bounded by
 ${\rm const.}\times (\epsilon N) \epsilon^{1/2}$.

 Hence, by a union bound, the probability that at least one variable $i\in T$ is free is at most
 ${\rm const.} \times |T| (\epsilon N) \epsilon^{1/2}=c N^2 \epsilon^{5/2}$, for some constant $c>0$.

  Now, if no variable in $T$ is free, then $H^{eff}$ has exactly $1$ local minimum by lemma \ref{specialcase}.
  There are $2^{N-|T|}$ assignments to variables in $\overline T$, and we have established that at most
  $cN^2 \epsilon^{5/2} 2^{N-|T|}$ such assignments have more than one local minimum.
  Hence, there are at most
  \be
  2^N\Bigl( 2^{-|T|} + cN^2\epsilon^{5/2}\Bigr)
  \leq 2^N \Bigl(2^{-((1/2)\epsilon N -1)}+cN^2\epsilon^{5/2}\Bigr)
  \ee
  local minima for $H$.
  Choosing $\epsilon=\log(N)/N$, we find that the above equation is bounded by
  $$2^N \Bigl( 2/\sqrt{N}+{\rm polylog}(N)/\sqrt{N}\Bigr),$$ where
  the polylog is bounded by a constant times $\log(N)^{5/2}$.
\end{proof}
\end{theorem}
\section{Sparse Case}
\label{spc}
In this section, we give an upper bound on local minima for the sparse case.
Given an Ising instance, define a graph $G$ whose vertices correspond to variables, with an edge between two variables, $i,j$, if $J_{ij}\neq 0$.
Let $V$ be the set of vertices.
The degree of a vertex is defined as usual in graph theory; it is the number of edges attached to that vertex.

We will prove
\begin{theorem}
\label{badcase}
There are constants $\kb,\lb>0$ such that the following holds.
Consider an Ising instance and define the graph $G$ as above.  Suppose that $G$ has average vertex degree bounded by $d$.
Then, there are at most
$2^{N\cdot (1-\kb \log(\lb d)/d)}$
local minima.  Further, there is a deterministic algorithm taking polynomial space and time 
$\sO(2^{N\cdot (1-\kb \log(\lb d)/d)})$
which finds the assignment which minimizes $H$.
\end{theorem}

We prove this theorem by proving a similar bound in the case of bounded maximum degree:
\begin{theorem}
\label{bdcase}
There are constants $\kb,\lb''>0$ such that the following holds.
Consider an Ising instance and define the graph $G$ as above.  Suppose that $G$ has maximum vertex degree bounded by $d$.
Then, there are at most
$2^{N\cdot (1-\kb' \log(\lb' d)/d)}$
local minima.  Further, there is a deterministic algorithm taking polynomial space and time 
$\sO(2^{N\cdot (1-\kb' \log(\lb' d)/d)})$
which finds the assignment which minimizes $H$.
\end{theorem}

{\it Proof of theorem \ref{badcase} assuming theorem \ref{bdcase}:} let $W$ be the set of variables with degree at most $2d$.  Since $G$ has average vertex degree $d$, $|W| \geq N/2$.
For each of the $2^{N-|W|}$ assignments to variables in $\overline W$, we construct an effective Hamiltonian for variables in $W$.
Applying theorem \ref{bdcase} to this effective Hamiltonian, shows that there are at most $2^{|W|\cdot (1-\kb' \log(2\lb 'd)/(2d))}$ local minima of this Hamiltonian and so there are at most
$2^{N-|W|}2^{|W|\cdot (1-\kb' \log(2\lb' d)/(2d))}\leq 2^{N\cdot (1-\kb' \log(2\lb' d)/(4d))}$ local minima of $H$.  Similarly, we can minimize 
$H$ by iterating over assignments to variables in $\overline W$ and then minimizing the effective Hamiltonian using the algorithm of theorem \ref{bdcase}.  So, theorem \ref{badcase} follows with $\kb=\kb'/4$.

So, we now focus on proving theorem \ref{bdcase}.
First, in subsection \ref{Tsubsec}, we construct a set $T$ which is in some ways analogous to the set $T$ constructed in the non-sparse case above in that vertices in $T$ will have many edges with large $|J_{ij}|$ to vertices $j \not \in T$ and will have small $h^{max}_i\equiv \sum_{j\in T}|J_{ij}|$.  Then, in subsection \ref{minalg}, we complete the proof of the theorem, by showing that
for a random assignment to vertices in $\overline T$, the effective Hamiltonian $H^{eff}$ for vertices in $T$ will have many vertices fixed.
The number of local minima  of $H^{eff}$ will be bounded by $2^{|F|}$ where $F$ is the set of free variables, and we will bound the sum of this quantity over all assignments to vertices in $\overline T$.  We will refer to this sum as a ``partition function", $Z$.
Then, the algorithm of theorem \ref{bdcase} will simply be: (1) construct $T$ (2) iterate over assignments to variables in $\overline T$.  For each assignment, compute $H^{eff}$ (this can be done in polynomial time) and then compute the set $F$ of free variables in $T$.  An optimal assignment to the fixed variables (those in $T\setminus F$) can be computed in linear time given $H^{eff}$, and then one can define a new effective Hamiltonian for the variables in $F$, taking the assignment to the variables in $T\setminus F$ and the variables in $\overline T=V\setminus T$ as given.  Finally, this new effective Hamiltonian can have its minimum found by iterating over all assignments to variables in $F$.  Since there are $2^{|F|}$ such assignments, the total run time is equal to the sum over all assignments to variables 
in $\overline T$ of $2^{|F|}$ times a polynomial; i.e., it is equal to $Z$ times a polynomial.  The polynomial factor is the time required to
compute the effective Hamiltonians and find the set $T$ and other sets.

\subsection{Construction of set $T$}
\label{Tsubsec}
We will give first a randomized construction of $T$, and then use that to give a deterministic algorithm to find $T$.  We will have $|T|=\Theta(\epsilon N)$ where we later pick $\epsilon=\log(d)/d$.
The run time of the algorithm will be exponential in $N$, but for small $\epsilon$ this runtime is small compared to the upper bounds on the runtime of the algorithm in theorem \ref{bdcase}.

It is important to understand that the particular choice of $\epsilon$ does not matter too much.  We have picked an optimal value (up to constants) for the proof as wlll be clear later.  However, to give a rough idea of the appropriate value of $\epsilon$: first, we need to pick $\epsilon$ at least ${\rm const.}\times \log(d)/d$, as otherwise, even if all variables in $T$ were fixed with probability $1$, we would not obtain a good bound on the number of local minima.  Second, we can actually pick $\epsilon$ significantly larger than $\log(d)/d$; we could have for example picked $\epsilon=d^{-\alpha}$ for any exponent $\alpha>1/2$ and we would still have a meaningful bound (though not quite as tight.  The point is that we will need $d_{T,\overline T}$ (as defined in the lemma) large enough that
$\sqrt{d_{T,\overline T}} >> d_T$ so that the interactions within the set $T$ (which at worst case have strength $\sim d_T$) are small compared to the interactions between a variable in $T$ and a variable in $\overline T$ (which on average have strength $\sim \sqrt{d_{T,\overline T}}$).

\begin{lemma}
\label{Tfind}
Assume the conditions of theorem \ref{bdcase} hold.
For all sufficiently small $\epsilon$,
there is a set $T \subset V$ with $|T|=\Theta(\epsilon N)$, such that the following properties hold.

Define the graph $G$ as above. Define a bipartite graph $G_{T,\overline T}$ containing only the edges between vertices $i \in T$ and $j\in \overline T$.
Define a graph $G_{T}$ which is the induced subgraph of $G$ containing only vertices $i\in T$.

Then, first, for every $i \in T$, the degree of that vertex in $G_{T}$ is at most $d_T$ where $$d_T\equiv 99\epsilon d.$$
Second, for every $i\in T$, the number of $j \in \overline T$ such that $|J_{ij}| \geq {\rm max}_{k \in T} |J_{ik}|$ is at least $d_{T,\overline T}$ where
$$d_{T,\overline T}=\lfloor (1/99)\epsilon^{-1} \rfloor.$$
For each $i$, if the degree of the vertex $i$ in $G_T$ is nonzero, then
we pick $d_{T,\overline T}$ edges in $G_{T,\overline T}$ which connect $i$ to $j$ such that $|J_{ij}| \geq {\rm max}_{k \in T} |J_{ik}|$ and we call these
``strong edges".
Then, third, for every $i\in T$, the sum over first neighbors of $i$ in $G_{T,\overline T}$ of the number of strong edges attached to that first neighbor is at most
$\Delta$ with
$$\Delta=99 d.$$

Further, such a set $T$ can be constructed by a deterministic algorithm
taking time ${\rm poly}(N) {N \choose |T|}$.  This time is $\sO(c_0^N)$, where $c_0$ tends to $1$ as $|T|/N$ tends to $0$.
\begin{proof}
Let $T_0$ be a randomly chosen subset of $V$ where we independently choose for each vertex whether or not it is in $T_0$, choosing it to be in $T_0$ with probability $\epsilon$.  With high probability, $|T_0|\geq (1-o(1)) \epsilon N$.

 We will test various properties of the vertices, labeling certain vertices as ``good" or ``bad" depending on whether or not they obey these properties.
Define a bipartite graph $G_{T_0,\overline T_0}$ containing only the edges between vertices $i \in T_0$ and $j\not \in T_0$.
Define a graph $G_{T_0}$ which is the induced subgraph of $G$ containing only vertices $i\in T_0$.

Every  $i\in T_0$ which has degree $0$ in $G_{T_0}$ is labelled as good.
Every $i\in T_0$ which has nonzero degree in $G_{T_0}$ is labelled as
``good" if the following three properties hold, and otherwise we label $i$ as bad.
First, the degree of that vertex in $G_{T_0}$ is at most $d_T$.
Second, the number of distinct $j\not \in T_0$ such that $|J_{ij}| \geq {\rm max}_{k \in T_0} |J_{ik}|$ is at least $d_{T,\overline T}$.
Assuming $i$ obeys these two criteria, we then randomly choose $\lfloor d_{T,\overline T} \rfloor$ of these $j$ and call the edge connecting $i,j$ a strong edge.
Then,
third, for every $i\in T_0$,
the sum over first neighbors of $i$ in $G_{T,\overline T}$ of the number of strong edges attached to that first neighbor is at most
$\Delta$.
What we will show is a random vertex in $T_0$ has at least some constant positive probability of being good.  Hence, there is a choice of $T_0$ such that at least a constant fraction of vertices in $T_0$ are good.  We then set $T$ to be set of good vertices in $T_0$, and the desired properties will follow.

We upper bound the probability that a random vertex $i\in T_0$ does not obey each of the three properties above, using three separate first moment bounds.  We then apply a union bound to upper bound the probability that the vertex is bad.

The expected degree of $i$ in $G_{T_0}$ is at most $\epsilon d$, so the probability that the vertex does not obey the first property above is at most $1/99$.

The probability that the number of distinct $j\not \in T$ such that $|J_{ij}| \geq {\rm max}_{k \in T} |J_{ik}|$ is at least $d_{T,\overline T}$ can be bounded similarly to the non-sparse case: construct a set of $j \not \in T$ containing $d_{T,\overline T}$ elements such that $|J_{ij}|$ is at least as large for all $j$ in this set as $|J_{ik}|$ is for all $k$ not in this set.  Note that if $i$ has degree sufficiently small, then it may be necessary to include some $j$ such that $J_{ij}=0$.
Then, the average number of $j$ in this set which are in $T_0$ is $\epsilon d_{T,\overline T}=1/99$, so there is a probability at most $1/99$ that $i$ does not obey the second property.

Before giving a detailed proof of the third property, let us give a heuristic estimate: There are $\Theta(\epsilon N)$ variables in $T$, each with $\Theta(\epsilon^{-1})$ strong edges, so that there are $\Theta(N)$ strong edges in total, so the average number of strong edges attached to each vertex in $\overline T$ is $\Theta(1)$.  Each $i \in T$ has at most $d$ edges to vertices in $\overline T$, so on average one might guess that there are $\Theta(d)$ edges attached to those vertices.  To give a proof, though, we need to consider correlations more carefully, as we now do.  Note, however, that this heuristic (and the proof below) both show that the value of $\Delta$ does not depend on the $\epsilon$ that we have chosen; this is one of the reasons why the proof of the theorem would work even if we had chosen a larger value of $\epsilon$, as we noted above this lemma.

To show the third property, we first bound the number of triples $i,j,k$ with $i,k \in T_0$ and $j \not \in T_0$ with $i,j$ connected
by an edge and $j,k$ connected by a strong edge, and $k$ good.
First consider the triples with $i,k$ neighbors.  There are at most $|T_0|$ choices of $k$ and for each $k$ there are at most $d_T$ neighbors $i\in T_0$ and at most $d_{T,\overline T}$ strong edges in total, so there are at most $|T_0| d_T d_{T,\overline T} \leq |T_0| d$ such triples.
Now consider the triples with $i,k$ not neighbors.  There are at most $|T_0| d_{T,\overline T}$ choices of $j,k$ and there are at most
$|T_0| d_{T,\overline T} d$ vertices $i$ which neighbor $j$.
Since $k$ is not a neighbor of $i$, the set of strong edges attached to $k$ is independent of whether or not such $i$ is in $T_0$, and 
hence there are on average at most $|T_0| d_{T,\overline T} d \epsilon\leq |T_0| d/99$ such triples.  Hence, there are on average at most $|T_0| d(1+1/99)$ triples in total.
Hence, for random $i\in T_0$, the expected number of such $j,k$ is at most $d(1+1/99)$.  Hence, the probability that the vertex $i$ does not obey the third property is at most $(1/99)(1+1/99)$.

So, by a union bound, $i$ is bad with probability at most $1/99+1/99+(1/99)/(1/99+1)<4/99$ and so is good with probability $>95/99$.

So, for a random choice of $T_0$, the average number of good variables is at least $(95/99) |T_0|$ and so there must be some choice of $T_0$
such that there are at least $(95/99)(1-o(1)) \epsilon N$ good variables.  Choose $T$ to be the set of good variables for that choice of $T_0$.

This gives a randomized construction of $T$.  However, a set $T$ with these properties can be constructed using a deterministic algorithm simply by iterating
over all ${|W| \choose |T|} \leq {N \choose |T|}$ choices of $T$ and checking each choice to see whether it has these properties.
\end{proof}
\end{lemma}

\subsection{Bound on Minima and Algorithm}
\label{minalg}
We consider an assignment to variables in $V\setminus T$.  We denote this assignment $A$.  
We will choose
$$\epsilon=\log(d)/d$$ in the construction of the set $T$.

We define $H^{eff}$ as before:
\be
H^{eff}=\frac{1}{4}\sum_{i \in T} h^{eff}_i S_i+
\frac{1}{4}\sum_{i<j, i\in T, j\in T} J_{ij} S_i S_j,
\ee
where
\be
h^{eff}_i=h_i+\sum_{j \not \in T} J_{ij} S_j.
\ee
We define a ``partition function" $Z$ (to borrow the language of statistical physics) to be the sum, over all assignments $A$ to variables in $V\setminus T$,
of $2^{|F|}$.  We will estimate $Z$.  Then, as explained at the start of the section, this gives the desired bound on the number of local minima and gives the desired algorithm.

We will estimate the number of fixed variables by constructing a sequence of variables $i_1,i_2,\ldots \in T$, where every variable in $T$ appears in the sequence exactly once.  As we construct this sequence,
we estimate the probability that a given variable $i_a$ is fixed or free, where the probability is over random assignments to variables in $V\setminus T$.
Thus, the partition function $Z$ is equal to $2^N$ times the sum over all possible sequences of events (i.e., for each $i_a$, whether $i_a$ is fixed or not, so that there are $2^{|T|}$ possible sequences of events) of the probability of that sequence of events times $2^{|F|}$.

The probability that $i_1$ (the first variable in the sequence) is fixed
will be easy to estimate similarly to the sparse case: the terms $J_{ij} S_j$ will be uncorrelated random variables, so $h^{eff}_{i_1}$ is a sum of uncorrelated random
variables.  However, the event that $i_a$ is fixed for $a>1$ will be correlated with the previous events of whether or not $i_1,i_2,\ldots,i_{a-1}$ are fixed.

To keep track of these correlations, we introduce another sequence of sets, $R(a)\subset V\setminus T$; the set $R(a)$ is a set of variables $j$ for $j \not \in T$, that have been ``revealed" as explained later.  We set $R(0)=\emptyset$.
During the construction the set $R$
will have additional variables added to it as follows.  If variable $i_a$ is free, then $R(a)$ is equal to $R(a-1)$ union the set of $j$ such that $J_{i_a j}\neq 0$.  If variable $i_a$ is fixed, then $R(a)=R(a-1)$.

We explain later how we choose the $i_1,i_2,\ldots$.  The choice of $i_a$ will depend upon the previous events, i.e., on which $i_b$ for $b<a$ are fixed.

Consider the $a$-th step.  We wish to estimate the probability that $|h^{eff}_{i_a}| < h^{max}_0$.  We write
\be
h^{eff}_{i_a}=h^{eff,0}_{i_a}+h^{eff,1}_{i_a},
\ee
where
\be
h^{eff,0}_{i_a}=h_{i_a}+\sum_{j \in R(a-1)} J_{ij} S_j,
\ee
and
\be
h^{eff,1}_{i_a}=\sum_{j\in (V\setminus T)\setminus R(a-1)} J_{ij} S_j.
\ee
So, the probability that $|h^{eff}_{i_a}|< h^{eff}$ is equal to the probability that $|h^{eff,1}_{i_a}+h^{eff,0}| < h^{max}$.

Consider a given sequence of events, such as $i_1$ fixed, $i_2$ free, $i_3$ free, $i_4$ fixed.
We have
\begin{eqnarray} 
&& \Pr(i_1,i_4\in F; i_2,i_3 \not \in F)
\\ \nonumber
&=& \Pr(i_4 \in F|i_1 \in F, i_2,i_3\not \in F) \Pr(i_3 \not \in F|  i_1 \in F,i_2 \not \in F) \Pr(i_2 \not \in F|i_1 \in F) \\ \nonumber
&\leq & \Pr(i_4 \in F| i_1 \in F) \Pr(i_1 \in F).
\end{eqnarray}
In general, we can apply this inequality to any sequence of events: the probability that the set $F$ contains exactly the variables $i_{a_1},i_{a_2},i_{a_3},\ldots$ for $a_1<a_2<a_3\ldots $  is bounded by the product of conditional probabilities, $\Pr(i_{a_3} \in F|i_{a_1},i_{a_2}\in F) \Pr(i_{a_2} \in F|i_{a_1}\in F)  \Pr(i_{a_1} \in F) $.
This inequality is behind the usage of the term ``revealed" above: by computing just this product of conditional probabilities, where the only events conditioned are events where variables are found to be in $F$ and we never condition on an event that a variable is not in $F$, we can treat all the terms $J_{i_aj}$ for $j$ that have not been revealed as independent random events.  

To compute a probability such as $\Pr(i_{a_k} \in F|i_{a_1}, \ldots, i_{a_{k-1}}\in F)$, we compute the probability that 
$|h^{eff,1}_{i_{a_k}}+h^{eff,0}_{i_{a_k}}| \leq h^{max}$.  The random variable $h^{eff,0}_{i_{a_k}}$ may be correlated with the event that
$i_{a_1}, \ldots, i_{a_{k-1}}\in F$ in some complicated way, and thus conditioning on this event may give some complicated distribution to this random variable.  However, the random variable $h^{eff,1}_{i_{a_k}}$ is uncorrelated with the event that $i_{a_1}, \ldots, i_{a_{k-1}}\in F$.
We have 
\begin{eqnarray}
&&
\Pr(i_{a_k} \in F|i_{a_1}, \ldots, i_{a_{k-1}}\in F)
\\ \nonumber
&=&
\sum_{h} \Pr(h^{eff,0}_{i_{a_k}}=h|i_{a_1}, \ldots, i_{a_{k-1}}\in F) \Pr(|h^{eff,1}_{i_{a_k}}+h|\leq h^{max})
\\ \nonumber
&\leq & 
{\rm max}_h
\Pr(|h^{eff,1}_{i_{a_k}}+h|\leq h^{max}_{i_{a_k}}).
\end{eqnarray}
At this point, we use lemma \ref{lemmarandvar} as in the non-sparse case.
Let $a_{min}={\rm max}_{j \in T} |J_{i_{a_k}j}|$, so
 $h^{max}_{i_{a_k}}\leq d_T a_{min}$.
For any $i$, let 
$d^{unr}_{i}(a)$ denote the number of strong edges connecting $i$ to vertices $j \not \in R(a-1)$.  That is,
it is the number of
distinct $j \in (V\setminus T)\setminus R(a-1)$ such that $|J_{ij}| \geq {\rm max}_{k \in T} |J_{ik}|$.  The suffix ``unr" is short for ``unrevealed".
Then, by lemma \ref{lemmarandvar},
${\rm max}_h \Pr(|h^{eff,1}_{i_{a_k}}+h|\leq h^{max}_{i_{a_k}}) \leq {\rm const.} \times d_T/\sqrt{d^{unr}_{i_{a_k}}(a)}$.
We next lower bound $d^{unr}_{i_{a_k}}(a)$.

Let $T(a)= T \setminus \{i_1,i_2,\ldots,i_{a-1}\}$ so that $|T(a)|=|T|-(a-1)$.
Let $E^{unr}(a)$ equal the sum of $d^{unr}_{i}(a)$ over $i \in T(a)$.
The average, over $i\in T(a)$, of $d^{unr}_i(a)$ is equal to $E^{unr}(a)/(|T|-(a-1))$.
Choosing $i_a$ to be a variable in $T(a)$ which maximizes $d^{unr}_{i_{a}}$,
we can ensure that
\be
d^{unr}_{i_a}(a) \geq E_{unr}(a)/(|T|-(a-1)).
\ee
Let $f(a)$ denote the number of variables $i_1,\ldots,i_a$ which are free.
So,
\be
E_{unr}(a)\geq (|T|-(a-1)) d_{T,\overline T}-\Delta f(a-1).
\ee
Hence,
\be
d^{unr}_{i_a}(a) \geq d_{T,\overline T}\Bigl(1-\frac{\Delta}{d_{T,\overline T}} \frac{f(a-1)}{|T|-(a-1)}\Bigr).
\ee
Hence,
\be
\label{big}
f(a-1)\leq \frac{1}{2} (|T|-(a-1)) \frac{d_{T,\overline T}}{\Delta} \quad \longrightarrow \quad d^{unr}_{i_a}(a)\geq \frac{d_{T,\overline T}}{2}.
\ee

Note that the left-hand side of the inequality on the left-hand side of the implication in Eq.~(\ref{big}) is a non-decreasing function of $a$ while the right-hand side of that inequality is a decreasing function of $a$, so if the inequality fails to hold for some given $a$, then it also fails for all larger $a$.
Given a sequence of events of whether or not $i_1,i_2,\ldots,i_a$ are in $F$, we say that the sequence ``terminates at $a$" if $f(a-1)\leq \frac{1}{2} (|T|-(a-1)) \frac{\Delta}{d_{T,\overline T}}$ and
$f(a)> \frac{1}{2} (|T|-a) \frac{\Delta}{d_{T,\overline T}}$.
We can upper bound $Z$ by summing over sequences of events up to the step $a$ at which the sequence terminates and then using as an upper bound the assumption that after that point, all variables $i_b$ for $b>a$ are free with probability $1$.
Before the sequence terminates, we have 
${\rm max}_h \Pr(|h^{eff,1}_{i_{a_k}}+h|\leq h^{max}_{i_{a_k}}) \leq {\rm const.} \times d_T/\sqrt{d^{unr}_{i_{a_k}}(a)})\leq  {\rm const.} \times d_T/\sqrt{d_{T,\overline T}} $.
So,
\begin{eqnarray}
\label{term}
&&Z\\ \nonumber 
 & \leq & 2^{N} \sum_{a}  \sum_{\rm events}^{\rm seq. \, terminates \, at \, a} 2^{|F|-|T|} \prod_{b \leq a, {\rm s.t.}\, i_b \in F}(  {\rm const.} \times d_T/\sqrt{d_{T,\overline T}})\\ \nonumber
 & \leq & 2^{N} \sum_{a}  \sum_{\rm events}^{\rm seq. \, terminates \, at \, a}2^{-a} \prod_{b \leq a, {\rm s.t.}\, i_b \in F}2 \cdot (  {\rm const.} \times d_T/\sqrt{d_{T,\overline T}})\\ \nonumber
& = & 2^{N} \sum_{a}  \sum_{\rm events}^{\rm seq. \, terminates \, at \, a}2^{-a} \Bigl({\rm const.} \times d_T/\sqrt{d_{T,\overline T}}\Bigr)^{f(a)},
\end{eqnarray}
where the sum is over sequences of events terminating at the given $a$.
The factor in parenthesis on the first line is the upper bound on the probability that $i_b$ is free, conditioned on previous variables being free; the factor is bounded by $1$ because it is a probability.  
The factor $2^{-a}$ on the second line multiplied by the factor of $2$ for every $i_b\in F$ for $b\leq a$ is upper bounded by $2^{|F|-|T|}$.
On the last line, we absorbed the $2$ into the constant, so that the factor in parenthesis in the
last line is bounded by $2$.

In order for the sequence to terminate at $a$, we must have $f(a)> (1/2)(|T|-a)d_{T,\overline T}/\Delta$.  Thus, $a>|T|-2f(a)\Delta/d_{T,\overline T}$.
Thus,
\be
Z \leq 2^N \sum_{f(a)} \Bigl({\rm const.} \times d_T/\sqrt{d_{T,\overline T}}\Bigr)^{f(a)} \sum_{a>T-2f(a)\Delta/d_{T,\overline T}} {a \choose f(a)} 2^{-a}.
\ee
The factor ${a \choose f(a)}$ counts the number of sequences with the given $f(a)$.
The factor ${a \choose f(a)} 2^{-a}$ is exponentially small in $a$ unless $a \approx 2 f(a)$.  
We break the sum over $f(a)$ into two parts.  The first is a sum over $f(a)$ such that $2f(a)\Delta/d_{T,\overline T}\leq |T|/2$.  The second part is the sum over the remaining $f(a)$.
In the first sum, we always have $a\geq |T|/2 \geq 2 f(a)d_{T,\overline T}/\Delta$ so that the factor ${a \choose f(a)} 2^{-a}$ is exponentially small in $a$; we will have $d_{T,\overline T}<<\Delta$ so that in fact the
exponent is close to $1/2$.
Thus, the first sum is $O(c_1^{|T|})$ for a constant $c_1<1$ (the constant $c_1$ is slightly larger than $1/2$; the amount it is larger depends on $d_{T,\overline T}/\Delta$).  As for the second sum, each term is bounded by
$ \Bigl({\rm const.} \times d_T/\sqrt{d_{T,\overline T}}\Bigr)^{f(a)}$ where $f(a) \geq (|T|/2)d_{T,\overline T}/(2\Delta)$.
Since the number of terms in the sum is bounded by $|T|$,
the second sum is bounded by
$$|T| \Bigl({\rm const.} \times d_T/\sqrt{d_{T,\overline T}}\Bigr)^{(|T|/2)d_{T,\overline T}/(2\Delta)}.$$

Hence,
\be
\label{Zsbnd}
Z \leq 2^N \Bigl( O(c_1^{|T|})+|T| \Bigl({\rm const.} \times d_T/\sqrt{d_{T,\overline T}}\Bigr)^{(|T|/2)d_{T,\overline T}/(2\Delta)}\Bigr).
\ee

We have  $d_T/\sqrt{d_{T,\overline T}}=O(\log^{3/2}(d)/\sqrt{d})$ and $(|T|/2)d_{T,\overline T}/(2\Delta)=\Omega(N/d)$.
Hence,
\begin{eqnarray}
Z & \leq & 2^N \Bigl( O(c_1^{\epsilon N})+O(\log^{3/2}(d)/\sqrt{d})^{\Omega(N/d)} \Bigr)
\\ \nonumber
&=& 2^N O(2^{ -\Omega(N\log(d)/d}+2^{-\Omega(N \log(d)/d)})
\\ \nonumber
&=& 2^N 2^{-\Omega(N \log(d)/d)}.
\end{eqnarray}

The reader can now see why we have chosen $\epsilon$ as we did; it is so that both terms will be comparable in the above equation to get the optimal bound.
However, even if we had chosen $\epsilon$ larger ($\epsilon=d^{-\alpha}$ for $\alpha>1/2$), we would have still obtained a bound $Z \leq 2^N 2^{-\Omega(N \log(d)/d)}$.  The only way in which
the bound would be worse would be that the constant hidden by the $\Omega(\ldots)$ notation would be smaller.  The reason is that such a larger $\epsilon$ would still lead to $(|T|/2)d_{T,\overline T}/(2\Delta)=\Omega(N/d)$, but $d_T/\sqrt{d_{T,\overline T}}$ would be larger.

\section{Sum of Random Variables}
\label{secrandvar}
This section is devoted to the proof of the following lemma:
\begin{lemma}
\label{lemmarandvar2}
Let $\sigma_i$ for $i=1,\ldots,n$ be independent random variables, uniformly chosen $\pm 1$.
Let $\Sigma=\sum_i a_i \sigma_i$, with $|a_i|\geq 1$ for all $i$.
Let $\delta\geq 1$.
Then,
\be
{\rm max}_h {\rm Pr}(|\Sigma+h|\leq \delta) \leq {\rm const.} \times \frac{\delta}{\sqrt{n}}.
\ee
\begin{proof}
We have
\be {\rm max}_h {\rm Pr}(|\Sigma+h|\leq \delta)  \leq e^{1/2} {\rm max}_h E[\int \exp(-(\Sigma+h)^2/2\delta^2)],
\ee
where $E[...]$ denotes the expectation value over choices of $\sigma$.

Fourier transforming, we wish to evaluate
$${\rm const.} \times \delta \int {\rm d}k \exp(-k^2\delta^2/2) \exp(i k h) \prod_{i=1}^n \cos(a_i k),$$
for some numerical constant.
By a triangle inequality, this is bounded by
$${\rm const.} \times \frac{\delta}{\sqrt{2\pi}} \int {\rm d}k \exp(-k^2\delta^2/2) \prod_{i=1}^n |\cos(a_i k)|,$$
which is independent of $h$ so that we do not need to take a maximum.
We write 
\be
\label{intexpv}
\frac{\delta}{\sqrt{2\pi}} \int {\rm d}k \exp(-k^2\delta^2/2) \prod_{i=1}^n |\cos(a_i k)|=
E_\delta[\prod_{i=1}^n |\cos(a_i k)|],
\ee
where $E_\delta[\ldots]$ denotes an expectation value for a random choice of $k$ from the Gaussian 
$\frac{\delta}{\sqrt{2\pi}} {\rm d}k \exp(-k^2\delta^2/2)$.
This allows us to use the language of probability which will make certain arguments more clear.  For the remainder of the proof,
all probabilities and expectation values refer to expectation values with respect to this Gaussian distribution.

We define certain disjoint events.  The first will be the event where the product $\prod_{i=1}^n |\cos(a_i k)|$ is in the interval $(e^{-1},1]$.  The second is where the product is in the interval $(e^{-2},e^{-1}]$, and so on, so that in the $b$-th event, this product will be in the interval $(e^{-b},e^{1-b}]$.  In order for the $b$-th event to occur, it must be the case that
for at least half of the $i$, 
we have
$|\cos(a_i k)| \geq \exp(-2b/n)$.
To estimate the probability of the $b$-th event,
we claim (and we show in the next paragraph) that the probability (for any given $i$) that $|\cos(a_i k)|\geq e^{-2b/n}$ is bounded by ${\rm const.}\times \delta\sqrt{b/n}$.
Hence, the expected number of $i$ such that $|\cos(a_i k)|\geq e^{-2b/n}$ is bounded by ${\rm const.}\times \delta n\sqrt{b/n}$.
Hence, the probability that at least half of the $a_i$ have $|\cos(a_i k)|\geq e^{-2b/n}$ is bounded by $2\times{\rm const.}\times \delta \sqrt{b/n}$.
So, Eq.~(\ref{intexpv}) is bounded by
${\rm const.} \times \sum_{b=1}^{\infty} \exp^{1-b} \delta \sqrt{b/n}={\rm const.}\times \delta \sqrt{1/n}$. 

We finally show that the probability that $|\cos(a_i k)|\geq e^{-2b/n}$ is indeed bounded by ${\rm const.}\times \delta\sqrt{b/n}$.
If $\cos(a_ik)\geq e^{-2b/n}$, we have $\ln(|\cos(a_i k)|) \geq -2b/n$.  We have $\ln(|\cos(x)|)\leq {\rm max}_m -(x-\pi m)^2/2$ where the max is over integer $m$ (proof: it suffices to consider the case that $-\pi/2 < x < \pi/2$; on this interval, let $f(x)=\ln(|\cos(x)|)$ and let $g(x)=-x^2/2$; note that $f(0)=g(0)$ and $f'(0)=g'(0)$ where a prime denotes derivative and $f''(x)=-1/\cos^2(x) \leq g''(x)=-1$).
Hence, if $|\cos(a_ik)|\geq e^{-2b/n}$ then $a_i k$ is within distance $2\sqrt{b/n}$ of $m\pi$ for some integer $m$.
Hence, $k$ is within $2a_i^{-1} \sqrt{b/n}$  of $m\pi a_i^{-1}$.
So, the probability is bounded by
$$\sum_m \int_{m-2a_i^{-1}\sqrt{b/n}}^{m+2a_i^{-1}\sqrt{b/n}}\frac{\delta}{\sqrt{2\pi}} \exp(-k^2\delta^2/2) {\rm d}k.$$
For each choice of $m$, the integral is bounded by ${\rm const.} \times \delta a_i^{-1} \sqrt{b/n}$.
We distinguish two cases, either $\delta \geq a_i$ or $\delta<a_i$.  In the first case, the integral on the intervals with $m\neq 0$ decays exponentially in $m$, so that
the sum over $m$ is bounded by ${\rm const.} \times \delta a_i^{-1} \sqrt{b/n}$.
Using $|a_i|\geq 1$, this is bounded by ${\rm const.} \times \delta\sqrt{b/n}$.
In the second case, for $m>a_i/\delta$, the integral for each interval decays exponentially in $m\delta/a_i$, so that the probability is bounded by ${\rm const.} \times \sqrt{b/n}$, which is bounded by ${\rm const.}\times \delta\sqrt{b/n}$ since $\delta\geq 1$.
\end{proof}
\end{lemma}

\section{Combined Algorithm}
\label{combine}
Here we explain how to combine the ideas above with an algorithm from Ref.~\onlinecite{golo}.  The idea of is as follows.  First, the authors show the following lemma (lemma 4 in that reference, which we repeat here, slightly rephrased):
\begin{lemma}
\label{T1T2}
Let $G$ have average degree $d$ and $N$ vertices.  For any constant $0<\alpha<1$ and sufficiently large $d$, there exists two sets $T_1,T_2$, with $T_1 \cap T_2=\emptyset$ and $|T_1|,|T_2|=\alpha N \ln(d)/d$
such that there are no edges $(u,v)$ with $u\in T_1,v\in T_2$.
\begin{proof}
For a detailed proof, set Ref.~\onlinecite{golo}.  Here is a sketch of the proof: the proof is by the probabilistic method.  Choose $T_1$ at random.  Then, compute the probability that a vertex not in $T_1$ has no neighbors in $T_1$.  For the given $|T_1|$, this probability is large enough that the expected number of such vertices is greater than $\alpha N\ln(d)/d$.  Thus, there must be a choice of $T_1$ such that there are at least $\alpha N \ln(d)/d$ vertices with no neighbor in $T_1$.  Take this choice of $T_1$.
\end{proof}
\end{lemma}

Separately in Ref.~\onlinecite{golo} it is shown how to find these sets $T_1,T_2$ in time small compared to $\sO(2^{N\cdot(1-\alpha\ln(d)/d)})$.  One other way to do this is simply to iterate over all such sets.

Then, once these sets $T_1,T_2$ are found the algorithm is simply: iterate over all assignments to variables in $V\setminus (T_1 \cup T_2)$.  There are $2^{N\cdot(1-2\alpha\ln(d)/d)}$ such assignments.  
For each such assignment one can then find an optimal assignment for variables $T_1$ and $T_2$ separately (as no edges connect $T_1$ to $T_2$), and then combine the two assignments.  This takes time  $\sO(2^{N\cdot \alpha\ln(d)/d})$ for each assignment to variables in $V\setminus (T_1\cup T_2)$, giving the claimed total time.

We now show how to combine this idea with the method here in the case of an Ising instance for which all $J_{ij}$ are integers subject to a bound for all $i$ we have
\be
\label{Jmb}
\sum_j |J_{ij}| \leq J_{max},
\ee
for some $J_{max}$.  The results will be effective for $J_{max}$ sufficiently small compared to $d^{3/2}/\log^{3/2}(d)$.

The idea will be as follows.  Let $V_0=V \setminus (T_1 \cup T_2)$.  Then, find a  $T \subset V_0$ of size
$\Theta(\epsilon |V_0|)$ with $\epsilon=\log(d)/d$ such that all the conditions of lemma \ref{Tfind} hold for $T$ and such that additionally vertices in $T$ are only weakly coupled to vertices in $T_1 \cup T_2$, in a sense defined below.   Then, apply similar methods as before: iterate over all assignments to variables in $V_0\setminus T$; for most such assignments many of the variables in $T$ will be fixed independently of the choice of the variables in $T_1 \cup T_2$.  

We first need the following lemma which generalizes lemma \ref{Tfind}. This lemma will (like lemma \ref{Tfind}) assume bounded maximum degree, while lemma \ref{T1T2} assumed bounded average degree; however, the final theorem will only assume bounded average degree.
This lemma will allow $T_1,T_2$ to be arbitrary sets; we will not use any specific properties of them from lemma \ref{T1T2}.
This lemma modifies lemma \ref{Tfind} in two ways.  First, we have a lower bound on the degree of the vertices and we then require
that there be at least $d_{T,\overline T}$ strong edges connected to each $i\in T$ (in lemma \ref{Tfind}, vertices with degree $0$ in $G_T$ are allowed to have fewer than $d_{T,\overline T}$ edges in $G_{T,\overline T}$).  This change is done because, due to additional
interactions with $T_1,T_2$, we will need to have these additional edges to fix variables, while in the case of lemma \ref{Tfind}, every variable with degree $0$ in $G_T$ was automatically fixed.  The second change is the fourth condition on $T$ in the lemma, bounding interactions between variables in $T$ and those in $T_1 \cup T_2$.
\begin{lemma}
\label{Tfind2}
Consider an Ising instance and define the graph $G$ as above.  Suppose that $G$ has maximum vertex degree bounded by $d$ and all $J_{ij}$ are integers.
Let $T_1,T_2$ be two sets of vertices, with $V=V_0 \cup T_1 \cup T_2$ with $T_1,T_2,V_0$ disjoint.

Suppose further a randomly selected vertex in $V_0$ has degree at least $(1/99)\lfloor\epsilon^{-1}\rfloor$ with probability at least $98/99$.

For all sufficiently small $\epsilon$,
there is a set $T \subset V_0$ with $|T|=\Theta(\epsilon |V_0|)$, such that the following properties hold.

Define the graph $G$ as above. Define a bipartite graph $G_{T,\overline T}$ containing only the edges between vertices $i \in T$ and $j\not \in T$.
Define a graph $G_{T}$ which is the induced subgraph of $G$ containing only vertices $i\in T$.

Then, first, for every $i \in T$, the degree of that vertex in $G_{T}$ is at most $d_T$ where $$d_T\equiv 99\epsilon d.$$
Second, for every $i\in T$, the number of $j \in \overline T$ such that $|J_{ij}| \geq {\rm max}_{k \in T} |J_{ik}|$ is nonzero, and such that $J_{ij} \neq 0$, is at least $d_{T,\overline T}$ where
$$d_{T,\overline T}=\lfloor (1/99)\epsilon^{-1} \rfloor.$$
For each $i$,
we pick $d_{T,\overline T}$ edges in $G_{T,\overline T}$ which connect $i$ to $j$ such that $|J_{ij}| \geq {\rm max}_{k \in T} |J_{ik}|$ and such that $J_{ij} \neq 0$; we call these
``strong edges".
Then, third, for every $i\in T$, the sum over first neighbors of $i$ in $G_{T,\overline T}$ of the number of strong edges attached to that first neighbor is at most
$\Delta$ with
$$\Delta=99 d.$$
Fourth, for every $i\in T$, we have $\sum_{j \in (T_1 \cup T_2)} |J_{ij}| \leq 99 J_{max}(|T_1|+|T_2|)/|V_0|$.

Further, such a set $T$ can be constructed by a deterministic algorithm
taking time ${\rm poly}(N) {N \choose |T|}$.  This time is $\sO(c_0^N)$, where $c_0$ tends to $1$ as $|T|/N$ tends to $0$.
\begin{proof}
Choose a random subset $T\subset V$ as in lemma \ref{Tfind} and label vertices in that set as ``good" or ``bad" following the rules in the proof of lemma \ref{Tfind}.  

Additionally label a vertex as bad if it has degree less than $d_{T,\overline T}$.  By assumption, this occurs with probability at most $1/99$.  Thus, all good vertices have degree at least $d_{T,\overline T}$ and hence if they are labelled good they will have at least that many strong edges.

Additionally,
label a vertex $i$ as bad if
$$\sum_{j \in (T_1 \cup T_2)} |J_{ij}| \geq 99J_{max} (|T_1|+|T_2|)/N.$$

Note that $\sum_{i \in V_0} \sum_{j \in (T_1 \cup T_2)} |J_{ij}| \leq (|T_1|+|T_2|) J_{max} =|V_0| J_{max} (|T_1|+|T_2|)/|V_0| $.
So, the expectation for random $i \in V_0$ of $\sum_{j \in (T_1 \cup T_2)} |J_{ij}| $ is at most $J_{max} (|T_1|+|T_2|)/|V_0|$.
Hence, the probability that $\sum_{j \in (T_1 \cup T_2)} |J_{ij}| \geq 99 J_{max} (|T_1|+|T_2|)/|V_0|$ is bounded by $1/99$.

Using a similar union bound to lemma \ref{Tfind}, there is a choice of $T_0$ for which at least $(93/99)(1-o(1))$ of the variables in $T_0$ are good.  We take $T$ to be the set of
good variables for such a $T_0$.

This gives a randomized construction of $T$.  However, a set $T$ with these properties can be constructed using a deterministic algorithm simply by iterating
over all ${N \choose |T|}$ choices of $T$ and checking each choice to see whether it has these properties.
\end{proof}
\end{lemma}

Now we can prove:
\begin{theorem}
\label{badcasec}
There are constants $\kb,\lb>0$ such that the following holds.
Consider an Ising instance and define the graph $G$ as above.  Suppose that $G$ has average vertex degree bounded by $d$ and all $J_{ij}$ are integers and suppose that Eq.~\ref{Jmb} holds.
Then, for any $\alpha<1$ and all sufficiently large $d$, there is a deterministic algorithm taking polynomial space and time 
$\sO(2^{N\cdot (1-\alpha \ln(d)/d-\kb\log(P)/d)})$ 
which finds the assignment which minimizes $H$, where
\be
P=\lb\log^{3/2}(d)(1+ J_{max}/d)/\sqrt{d}.
\ee
\begin{proof}
We set $\epsilon=\log(d)/d$.
We consider two separate cases.  Either a randomly selected vertex in $G$ has degree at least $d_{T,\overline T}$ with probability at least $999/1000$, or the probability is smaller than this.  In the second case, where the probability is smaller than this, we use the following algorithm: iterate over all assignments to variables with degree larger than $d_{T,\overline T}$.  Then, for each such assignment, apply the algorithm of Ref.~\onlinecite{golo} to the effective Hamiltonian on the remaining variables.  The remaining variables are at least a $1/1000$ fraction of the variables but have degree at most $d_{T,\overline T}=O(d/\log(d))$, and so we obtain a time $\sO(2^{N\cdot (1-\Omega(\log^2(d)/d))})$.

The rest of the proof considers the first case.
Then, we use the following algorithm:
first, construct $T_1,T_2$ using lemma \ref{T1T2}.   Let $W_0\subset V_0$ be the set of variables in $V_0$ with degree at most $2d$ (this is done to reduce to the bounded maximum degree case; this reduction step is essentially the same as how we reduced the proof of theorem \ref{badcase} to that of theorem \ref{bdcase}).  For large enough $d$, the size of $T_1,T_2$ will be such that 
a randomly selected vertex in $V_0$ has degree at least $(1/99)\lfloor\epsilon^{-1}\rfloor$ with probability at least $98/99$, one of the conditions of lemma \ref{Tfind2}.

An assignment to variables in $V_0 \setminus W_0$ defines an effective 
Hamiltonian for variables in $W_0 \cup T_1 \cup T_2$.  Construct $T$ using lemma \ref{Tfind2} applied to this effective Hamiltonian with $\epsilon=\log(d)/d$, using $W_0$ as the set called $V_0$ in lemma \ref{Tfind2}; in fact, this construction of $T$ needs to be done only once and the same $T$ can be used for all assignments to variables in $V_0 \setminus W_0$.

Now, iterate over all assignments to variables in $V_0 \setminus W_0$.
For each such assignment,  define the effective 
Hamiltonian for variables in $W_0 \cup T_1 \cup T_2$ and then iterate over all variables in $W_0 \setminus T$.
For $i \in T$, we define
$h^{max}_i\equiv \sum_{j \in (T \cup T_1 \cup T_2)} |J_{i,j}|$.
We say a variable $i$ in $T$ is free if $h^{max}_i > h^{eff}_i$ and otherwise we say $i$ is fixed.
Let $F$ be the set of such free variables.
We fix all fixed variables (this can be done in polynomial time) in $T$, and then iterate over free variables.
For each such free assignment, we then optimize over assignments in $T_1,T_2$ separately; this can be done in time $\sO({\rm max}(2^{|T_1|},2^{|T_2|}))$.

Thus, combining the iteration over variables in $V_0 \setminus W_0$ and $W_0 \setminus T$ into a single iteration, the algorithm is:
\begin{itemize}
\item[1.] Construct $T_1,T_2$ using lemma \ref{T1T2}.

\item[2.] Construct $W_0$ and $T$ using lemma \ref{Tfind2}.

\item[3.] Iterate over all assignments to variables in $V_0 \setminus T$.  For each such assignment, fix all fixed variables, and then iterate over free 
variables.  Then, optimize over assignments to $T_1,T_2$ separately.
\end{itemize}

To bound the time, we bound a partition function $Z$, which we define to be the sum, for a fixed assignment to variables in $V_0 \setminus W_0$, of t $2^{|F|}$ over all assignments to variables in $W_0 \setminus T$.
From here, the proof essentially follows that of theorem \ref{bdcase}.
It is important to understand that the factor
$-\Omega(N \log(d)/d)$ in the exponent of theorem \ref{bdcase} arises as follows. 
As shown below Eq.~(\ref{Zsbnd}), we have
$d_T/\sqrt{d_{T,\overline T}}=O(\log(d)/\sqrt{d})$ and $(|T|/2)d_{T,\overline T}/(2\Delta)=\Omega(N/d)$.
Hence,
$\Bigl({\rm const.} \times d_T/\sqrt{d_{T,\overline T}}\Bigr)^{(|T|/2)d_{T,\overline T}/(2\Delta)}=O(2^{-\Omega(N \log(d)/d)})$.
The factor ${\rm const.} \times d_T/\sqrt{d_{T,\overline T}}$ was an upper bound on the probability that a variable was free.
In the present case, the upper bound on the probability that a variable is free will be
$${\rm const.} \times (d_T+ 99 J_{max} (|T_1|+|T_2|)/|V_0|)/\sqrt{d_{T,\overline T}}$$.
\end{proof}
\end{theorem}

{\it Acknowledgments---} I thank S. Aaronson for very useful discussions.


\begin{thebibliography}{99}
\bibitem{maxcsp} On the hypercube $\{0,1\}^N$ every local maximum must neighbor $N$ assignments which are not local maxima, and every assignment which is not a local maximum can neighbor at most $N$ local maxima.
\end{thebibliography}
\end{document}